\documentclass[a4paper,11pt]{article}
\pdfoutput=1 

\usepackage{jheppub} 

\usepackage[T1]{fontenc} 

\title{\boldmath Fine Tuning Problem of the Cosmological Constant in a Generalized Randall-Sundrum Model}

\author[a,b]{Guang-Zhen Kang}
\author[c]{De-Sheng Zhang}
\author[b]{Hong-Shi Zong}
\author[b]{and Jun Li}

\affiliation[a]{School of Science, Yangzhou Polytechnic Institute,\\199 Huayang West Road, Yangzhou, China}
\affiliation[b]{Department of Physics, Nanjing University,\\22 Hankou Road, Nanjing, China,}
\affiliation[c]{School of Science, Changzhou Institute of Technology,\\1 Wushan Road, Changzhou, China,}

\emailAdd{gzkang@nju.edu.cn}

\abstract{To solve the cosmological constant fine tuning problem,
we investigate a $(n+1)$-dimensional generalized Randall-Sundrum brane world scenario with two $(n-1)$-branes instead of two 3-branes.
Adopting an anisotropic metric ansatz, we obtain the positive effective cosmological constant $\Omega_{eff}$ of order $10^{-124}$
and only require a solution $\simeq50-80$.
Meanwhile, both the visible and hidden branes are stable because their tensions are positive.
Therefore, the fine tuning problem can be solved quite well.
Furthermore, the Hubble parameter $H_{1}(z)$ as a function of redshift $z$ is in good agreement with cosmic chronometers dataset.
The evolution of the universe naturally shifts from deceleration to acceleration.
This demonstrates that the evolution of the universe is intrinsically an extra-dimensional phenomenon.
It can be seen as a dynamic model of dark energy which is driven by the evolution of the extra dimensions on the brane.}

\begin{document}
\maketitle
\flushbottom

\section{Introduction}
\label{sec:intro}

The accelerated expansion of the universe has been discovered by analyzing distant and nearby supernovae in 1998 \cite{Riess,Perlmutter}.
This discovery indicated the return of cosmological constant which was proposed by Einstein, or termed as dark energy.
Data of last two decades indicate that the cosmological constant is a very tiny positive value of order $10^{-124}$ in Planck unit \cite{Perlmutter,Riess,Bennett,Halverson,Netterfield,Valent,Zhang,Verde,Moresco1,Moresco2,Moresco3,Ratsimbazafy,Stern,Simon}.
In Einstein's theory of general relativity, the cosmological constant coincides with the vacuum energy \cite{Peebles}.
Vacuum energy contains the contribution of vacuum fluctuations caused by quantum mechanics.
However, the theoretical expectations for the vacuum energy exceed the observed value by $124$ orders of magnitude \cite{Copeland}.
This gives rise to the cosmological constant fine tuning problem \cite{Amendola}.
Theorists have proposed numerous theoretical models \cite{Zlatev,Wetterich,Caldwell,Feng,Sotiriou,Dvali,Bousso} in order to solve this problem.
But these theories can only partially solve this problem or avoid the vacuum energy caused by quantum fluctuations.
It is still one of the central problems in theoretical physics and cosmology.

We have investigated some high-dimensional theory \cite{KK,NAH1,NAH2,Antoniadis,RS,Das,Das1,Antoniadis1,Sundrum,Lykken,Visinelli,Vagnozzi,Paul,Polchinski} in order to solve this problem better.
Among these theories, the Randall-Sundrum (RS) two-brane model \cite{RS} has attracted our attention.
In this model the hierarchy problem why there is such a large discrepancy between the electroweak scale/Higgs mass $M_{EW}\sim1$TeV and the Planck mass $M_{pl}\sim10^{16}$TeV is solved very well.
The warp factor $e^{-2kr\pi}$ originating from the background metric generates a large hierarchy but requires only $kr\approx10$.
This large hierarchy problem is somewhat similar to the cosmological constant fine tuning problem.
However, RS model has a negative brane tension which results in the instability of the visible brane.
Furthermore, zero cosmological constant on the visible 3-brane is not consistent with presently observed data \cite{Das,Koley}.
Therefore one solution of the fine tuning problem is to find a new mechanism in RS model that renders the effective cosmological constant a very small positive value compatible with the present cosmological observation.

In this paper, we show that this problem can be solved using the $(n+1)$-dimensional (-d) generalized RS braneworld model \cite{Das} with two $(n-1)$-branes (the visible and the hidden branes) instead of two 3-branes.
In this model, the induced cosmological constant $\Omega$ on the visible brane appears naturally.
For the negative induced cosmological constant,
there is a solution with positive visible and hidden brane tension \cite{Mitra,SC1,SC2,SC3,Banerjee}.
For $-\Omega$ of order $10^{-124}$,
we obtain the solutions corresponding to the magnitude of the warp factor.

In order to be consistent with observations,
the negative induced cosmological constant must be transformed into the positive effective cosmological constant.
We investigate the case that the scale factors on the visible brane evolve with different rates by adopting an anisotropic metric ansatz \cite{Middleton}.
The Hubble parameters are obtained by solving the Einstein field equation.
We obtain a positive effective cosmological constant $\Omega_{eff}$ consistent with observations of order $-\Omega$.
It indicates that cosmic acceleration is intrinsically an extra-dimensional phenomenon.
The Hubble parameter $H_{1}(z)$ as a function of redshift $z$ is consistent with most of cosmic chronometers dataset.
More precise cosmic chronometers dataset may verify the existence of extra dimensions on the visible brane.

\section{$(n+1)$-d Generalized Randall-Sundrum model}

We start with a $(n+1)$-d generalized RS braneworld model.
The total action $S_{n+1}$ is composed of the bulk action and
two $(n-1)$-brane actions:
\begin{eqnarray}
S_{bulk}&=&\int d^{n}xdy\sqrt{-G}(M^{n-1}_{n+1}R-\Lambda)\label{eq:S1},
\end{eqnarray}
\begin{eqnarray}
S_{vis}&=&\int d^{n}x\sqrt{-g_{vis}}(\mathcal{L}_{vis}-V_{vis}),
\end{eqnarray}
\begin{eqnarray}
S_{hid}&=&\int d^{n}x\sqrt{-g_{hid}}(\mathcal{L}_{hid}-V_{hid}),
\label{eq:S2}
\end{eqnarray}
where $\Lambda$ is a bulk cosmological constant of order $1$ in Planck unit (the vacuum energy caused by quantum mechanics),
$M_{n+1}$ denotes $(n+1)$-d fundamental mass scale,
$G_{AB}$ is the $(n+1)$-d metric tensor,
$R$ is the $(n+1)$-d Ricci scalar,
$\mathcal{L}_{i}$ and $V_{i}$ are the matter field Lagrangian and the tension of the $i-th$ brane, $i=hid$ or $vis$.
The total action $S_{n+1}$ is stationary with respect to arbitrary variations in $G_{AB}$ if and only if:

\begin{eqnarray}\label{eq:field}
R_{AB}-\frac{1}{2}G_{AB}R=\frac{1}{2M^{n-1}_{n+1}}\{-G_{AB}\Lambda+\sum_{i}[T^{i}_{AB}\delta(y-y_{i})-G_{ab}\delta_{A}^{a}\delta_{B}^{b}V_{i}\delta(y-y_{i})]\},
\end{eqnarray}
where Capital Latin $A,B$ indices run over all spacetime coordinate labels $0, 1, 2, \cdot\cdot\cdot, n$, Lowercase Latin $a,b=0,1,2,\cdot\cdot\cdot,n-1$,
$R_{AB}$ and $T^{i}_{AB}$ are the $(n+1)$-d Ricci and the energy-momentum tensors respectively,
$y_{i}$ represents the position of the $i$-th brane in the $(n+1)$-th coordinate.
The $(n+1)$-d energy-momentum tensor of an anisotropic perfect fluid is
$T^{iA}_{B}=diag[-\rho_{i},p_{i1},p_{i2},\cdot\cdot\cdot,p_{i(n-1)},0]$.
In this generalized RS model, the metric ansatz satisfying the Einstein filed equations Eq.~(\ref{eq:field}) is of the form:
\begin{eqnarray}\label{eq:ds2}
ds^{2}=G_{AB}dx^{A}dx^{B}=e^{-2A(y)}g_{ab}dx^{a}dx^{b}+r^{2}dy^{2},
\end{eqnarray}
where $g_{ab}$ is the $n$-d metric tensor, $e^{-2A(y)}$ is known as the warp factor,
$y$ is the extra dimensional coordinate of length $r$.
Combining the Eq.~(\ref{eq:field}) with Eq.~(\ref{eq:ds2}), each energy-momentum tensor is given by $T^{ia}_{b}=diag[-c_{i},c_{i},\cdot\cdot\cdot,c_{i}]$.
This is similar to the RS model in which each 3-brane Lagrangian separated out a constant vacuum energy \cite{RS}.
Then, the equations of $A''$ and $A'^{2}$ are given by:
\begin{eqnarray}
A''&=&\dfrac{2\Omega e^{2A}}{(n-1)(n-2)}+\frac{\sum_{i}\delta(y-y_{i})\mathcal{V}_{i}}{2M^{n-1}_{n+1}(n-1)},\label{eq:A1}\\
A'^{2}&=&\dfrac{2\Omega e^{2A}}{(n-1)(n-2)}+k^{2}\label{eq:A2}
\end{eqnarray}
where we redefine the brane tension $\mathcal{V}_{i}\equiv V_{i}-c_{i}$,
the constant $k\equiv\sqrt{-\Lambda/[M^{n-1}_{n+1}n(n-1)]}\simeq$ Planck mass,
the arbitrary constant $\Omega$ corresponds to the induced cosmological constant on the visible brane.
Meanwhile, the $n$-d Einstein field equations with this induced cosmological constant are given by:
\begin{equation}\label{eq:ndfield}
\widetilde{R}_{ab}-\frac{1}{2}g_{ab}\widetilde{R}=-\Omega g_{ab}.
\end{equation}
where $\widetilde{R}_{ab}$ and $\widetilde{R}$ are the $n$-d Ricci tensor and Ricci scalar, respectively.
The arbitrary constant $\Omega$ corresponds to the induced cosmological constant on the visible brane.
In this paper, we only consider the situation $\Omega<0$.
Since $\Omega>0$ always brings about the negative tension on the visible brane that results in instability.

For $\Omega<0$,
the warp factor satisfying Eqs.~(\ref{eq:field}) and (\ref{eq:ds2}) is obtained:
\begin{eqnarray}\label{eq:solution-}
A=-\ln[\omega\cosh(k|y|+c_{-})],
\end{eqnarray}
where $\omega^{2}\equiv-2\Omega/[(n-1)(n-2)k^{2}]$, $c_{-}=\ln[(1-\sqrt{1-\omega^{2}})/\omega]$  for considering the normalization at the extra dimensional fixed coordinate $y=0$.
Setting $e^{-A(r\pi)}\equiv10^{-m}$, $x\equiv\pi kr$, $\omega^{2}\equiv10^{-l}$ \cite{Das, Kang},
we get from Eq.~(\ref{eq:solution-}):
\begin{eqnarray}\label{eq:x2}
e^{-x}=\frac{10^{-m}}{2}[1\pm\sqrt{1-10^{-(l-2m)}}],
\end{eqnarray}
where the induce cosmological constant have an upper bound ($l_{min}=2m$) to ensure that the value in the square root is greater than zero.
Eq.~(\ref{eq:x2}) have two solutions which give rise to the required warping.
For $(l-2m)\gg1$, the first solution corresponds to the visible brane tension
\begin{eqnarray}
\mathcal{V}_{vis}=-4(n-1)M^{n-1}_{n+1}k.
\end{eqnarray}
Here the visible brane is unstable because the visible brane tension is negative.
We do not consider this solution in the following.
The second solution $x=(l-m)\ln10+\ln4$, which is consistent with the Eq. (21) in Ref.~\cite{Das}.
Both the visible and the hidden brane are stable because both brane tensions are:
\begin{eqnarray}
\mathcal{V}_{vis}=\mathcal{V}_{hid}=4(n-1)M^{n-1}_{n+1}k>0.
\end{eqnarray}

\section{Fine tuning problem of the cosmological constant}

The $n$-d effective theory can be obtained by substituting Eq.~(\ref{eq:ds2}) into the bulk action Eq.~(\ref{eq:S1}).
We can obtain the scale of gravitational interactions from the curvature term:
\begin{eqnarray}\label{eq:Seff}
S_{eff}\supset\int d^{n}x\int_{-\pi}^{\pi}dy\sqrt{-g}M^{n-1}_{n+1}re^{-(n-2)A(kry)}\tilde{R}.
\end{eqnarray}
Substituting Eq.~(\ref{eq:solution-}) into Eq.~(\ref{eq:Seff}) and performing the integral of $y$, we can obtain a $n$-d effective action.
We find that $M_{n}$ depends only weakly on $r$ in the large $kr$ limit when $\omega^{2}\ll e^{-kr\pi}$.
Thus, $M_{n}$ is given by:
\begin{eqnarray}\label{Mn}
M^{n-2}_{n}\simeq\frac{2M^{n-1}_{n+1}}{(n-2)k}[1-e^{-(n-2)kr\pi}].
\end{eqnarray}
The hierarchy problem cannot be solved in this $(n+1)$-d generalized RS braneworld model
because that we can not determine the physical mass by means of renormalization.

If $l\sim124$, namely the magnitude of induced cosmological constant is of order $-124$ in Planck unit,
we get the second solution $x\simeq160-250$ corresponds to $m\simeq53-16$ when $m$ satisfies the condition $l-2m\gg1$.
Different warp factors correspond to different solutions to ensure $l\sim124$.
The cosmological constant fine tuning problem can be partially solved in this mechanism only requiring $kr\simeq50-80$.

The remaining unsolved problem is how to transform negative cosmological constant into positive.
We choose an anisotropic metric ansatz of the form \cite{Middleton}:
\begin{eqnarray}
g_{ab}=diag[-1,a_{1}^{2}(t),a_{2}^{2}(t),a_{3}^{2}(t),\cdot\cdot\cdot,a_{n-1}^{2}(t)],
\end{eqnarray}
where $a_{i}$ is the scale factor.
First, we investigate the case that the scale factors on the visible brane evolve with two different rates.
With the negative induced cosmological constant $\Omega\sim-10^{-124}$,
we obtain the solutions of Eq.~(\ref{eq:ndfield}):
\begin{eqnarray}
H_{1}&=&-\frac{\eta_{1}}{n-1}\tan\alpha+\eta_{2}\sec\alpha\label{eq:2h1},\\
H_{2}&=&-\frac{\eta_{1}}{n-1}\tan\alpha-\eta_{3}\sec\alpha\label{eq:2h2},
\end{eqnarray}
where the Hubble parameter $H\equiv\dot{a}/a$, $\alpha=\eta_{1}t+\theta_{0}$,
$\theta_{0}$ is the initial phase angle which is determined by the scale of the formation of the brane.
In other words, the value of $\theta_{0}$ must ensure that the presently observed three dimensional (3D) scale can be found.
The terms $\eta_{1}$, $\eta_{2}$ and $\eta_{3}$ are respectively:
\begin{eqnarray}
\eta_{1}=\sqrt{\frac{-2(n-1)\Omega}{n-2}}, \eta_{2}=\sqrt{\frac{-2\Omega n_{2}}{(n-1)n_{1}}}, \eta_{3}=\sqrt{\frac{-2\Omega n_{1}}{(n-1)n_{2}}},\nonumber
\end{eqnarray}
where $n_{1}$ and $n_{2}$ are the number of dimensions which evolve with two different rates, respectively.
We choose $n_{1}=3$ which is most in line with the presently observed 3D space universe.
From the time of the formation of the brane $t\equiv0$ to the present time ($t\sim10^{61}$ in planck unit),
it is easy to satisfy the condition that 3D scale factor is expanding all the time because $\eta_{1}$ is a very tiny value which is of order $\sqrt{-\Omega}\simeq10^{-62}$ in Planck unit.

\begin{figure}[tbp]
\centering 
\includegraphics[width=1.05\textwidth,trim=30 300 0 300,clip]{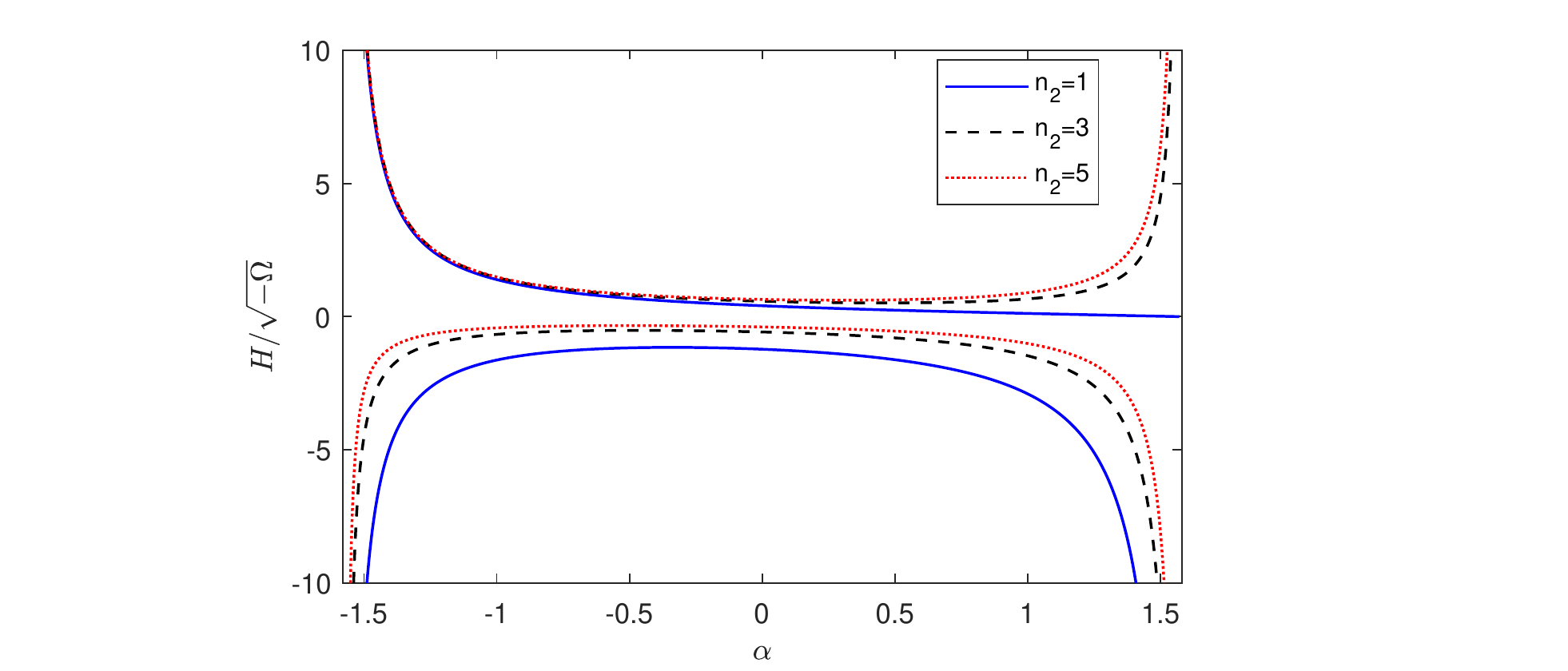}
\caption{\label{fig:1} (Color online). The Hubble parameters $H_{1}$ ($n_{1}=3$) and $H_{2}$.
Three different extra dimensions $n_{2}=1, 3, 5$ correspond to the solid (blue) curve, the dashed (black) curve and the dotted curve (red) respectively.}
\end{figure}

In Fig.~\ref{fig:1}, $H_{1}$ and $H_{2}$ as functions of $\alpha$ are shown.
The three curves in the upper half of Fig.~\ref{fig:1} correspond to $H_{1}$ at $n_{2}=1, 3, 5$ respectively.
It is shown that $H_{1}$ has a lower bound as long as $n_{2}\neq1$:
\begin{eqnarray}
H_{1min}=\sqrt{\eta_{2}^{2}-\eta_{1}^{2}/(n-1)^{2}},
\end{eqnarray}
where $\alpha_{min}=\arcsin\{\sqrt{n_{1}/[n_{2}(n_{1}+n_{2}-1)]}\}$.
Here we can see that the number of extra dimensions on the brane can not be one,
otherwise the time-independent positive effective cosmological constant cannot be obtained \cite{Kang}.
In the $n_{2}\rightarrow\infty$ limit, the lower bound of the Hubble parameter is $H_{1min}\simeq\sqrt{-6\Omega}/3$.
In the region near $\alpha_{min}$, $H_{1}$ is close to a constant of order $\sqrt{-\Omega}$.
Comparing with 3D Friedmann equation with $K=0$,
we obtain the effective cosmological constant is of order $-\Omega$
because in the region near $\alpha_{min}$ we have:
\begin{eqnarray}
H_{1}^{2}\sim-\Omega\equiv\Omega_{eff}>0.
\end{eqnarray}
This is an important result.
It tells us that the negative induced cosmological constant $\Omega$ can be transformed into the positive effective cosmological constant $\Omega_{eff}$ because of the evolution of the extra dimensions.
Cosmic acceleration is intrinsically an extra-dimensional phenomenon,
and the cosmological constant fine tuning problem can be solved by this extra-dimensional evolution.

\begin{figure}[tbp]
\centering 
\includegraphics[width=1.1\textwidth,trim=70 300 0 300,clip]{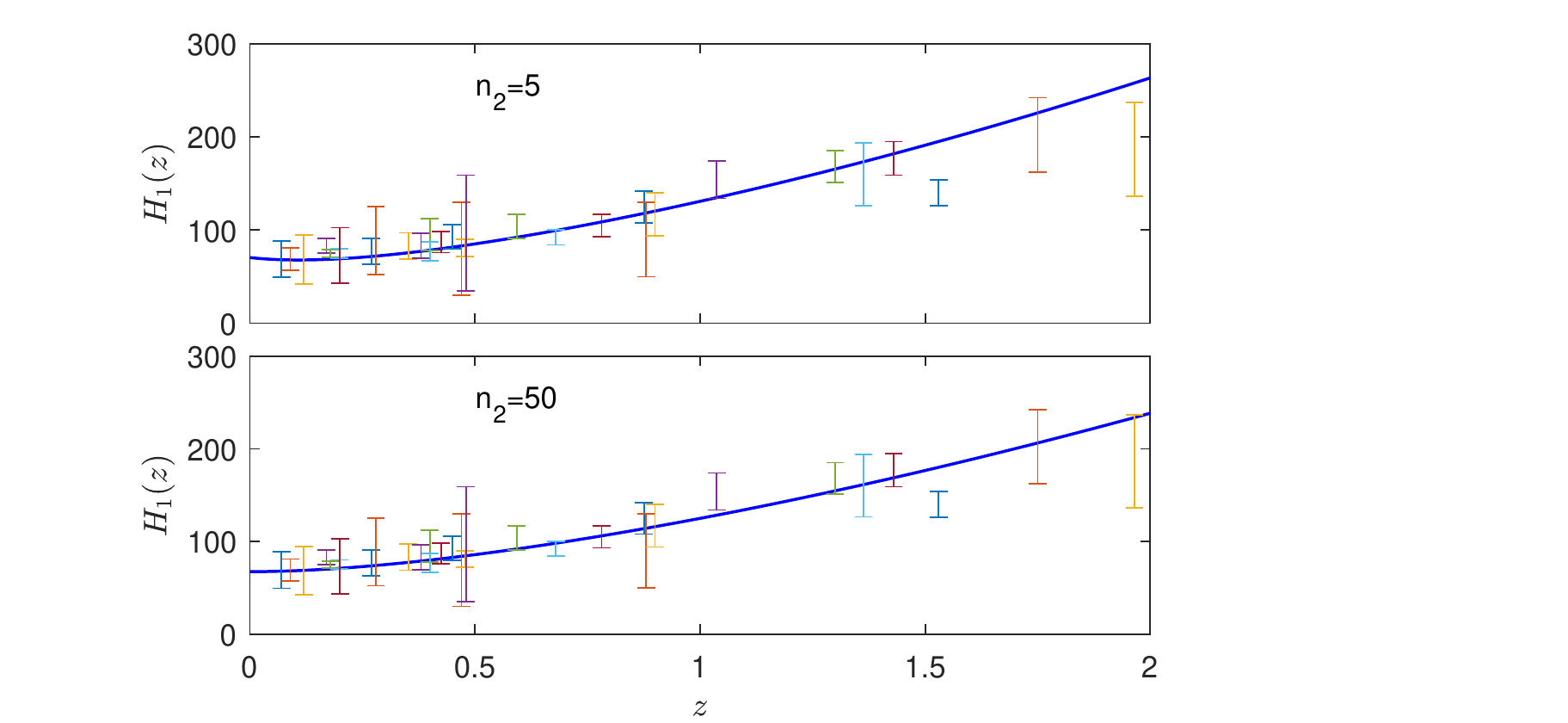}
\caption{\label{fig:2} (Color online). Cosmic chronometers dataset and $H_{1}(z)$ in [km/s/Mpc].
$H_{1}(z)$ ($n_{2}=5$ and $n_{2}=50$) as a function of redshift $z$ are compared with cosmic chronometers dataset extracted
from \cite{Valent,Zhang,Verde,Moresco1,Moresco2,Moresco3,Ratsimbazafy,Stern,Simon}.}
\end{figure}

From Eqs.~(\ref{eq:2h1}) and (\ref{eq:2h2}) we can get the scale factors $a_{1}$ and $a_{2}$.
Further, we obtain the volume of the visible brane:
\begin{eqnarray}
V_{b}=a_{1}^{n_{1}}a_{2}^{n_{2}}=a_{10}^{n_{1}}a_{20}^{n_{2}},
\end{eqnarray}
where $a_{10}$ and $a_{20}$ are the scale factors when the brane forms.
We choose $a_{10}=a_{20}$ because there is no reason to make them different when the brane has not evolved yet.
If the scale of the brane just formed is of order $10^{35}$ in planck unit ($\sim1m$),
in order to form the presently observed 3D scale of order $10^{61}$,
we obtain the scale of extra dimensions is of order $10^{9}$ with $n_{2}=3$.
Therefore, the scale of extra dimensions can be much larger than Planck length,
which leads to the fact that physics are still valid in this model.
The three curves in the lower half of Fig.~\ref{fig:1} presents $H_{2}$ with $n_{2}=1, 3, 5$ respectively.
With the increase of $n_{2}$, the scale of the brane just formed can be smaller,
meanwhile it ensures that the extra dimensions will not be reduce to Planck length in present time.
$\theta_{0}$ is close to $-\pi/2$ as long as $a_{10}$ is tiny.
So in the region of $\theta_{0}+\pi/2\ll\eta_{1}t\ll\pi/2$ we obtain:
\begin{eqnarray}
H_{1}\simeq\frac{1}{n_{1}+n_{2}}[1+\sqrt{\frac{(n_{1}+n_{2}-1)n_{2}}{n_{1}}}]\frac{1}{t}.
\end{eqnarray}
At $n_{2}=1$, $H_{1}$ is about $1/2t$ which is as similar as the radiation dominating eras.
In the limit $n_{2}\rightarrow\infty$, $H_{1}\simeq\sqrt{3}/3t$.

The figure on the top in Fig.~\ref{fig:2},
$H_{1}(z)$ as a function of redshift $z$ with $n_{2}=5$ is shown.
In order to best fit cosmic chronometers dataset,
we take the present time $t=t_{0}$ which is the time when $\alpha=\alpha_{min}+0.25$.
Therefore, $H_{1}(z)$ has a minimum at $z\sim0.1$.
Between $z=0$ and $z\sim0.1$, $H_{1}(z)$ is slowly decreasing which is not against most of cosmic chronometers dataset.
When $z>0.1$, $H_{1}(z)$ is monotonically increasing which is in agreement with cosmic chronometers dataset except at $z=1.965$.
The Hubble parameter is slightly larger than cosmic chronometers dataset at this redshift.
The redshift $z$ from deceleration to accelerated expansion (namely, deceleration parameter $q\equiv-\ddot{a}a/\dot{a}^{2}=0$) is about $0.35$.
It is somewhat inconsistent with the observed value $z\approx0.5$.
If more precise cosmic chronometers dataset show that the Hubble parameter is in line with the above curve,
the existence of extra dimension can be verified.
The figure on the bottom in Fig.~\ref{fig:2},
$H_{1}(z)$ is a monotonic curve when $n_{2}=50$.
We take the present time $t=t_{0}$ which is the time when $\alpha=\alpha_{min}$.
This curve is in good agreement with cosmic chronometers dataset very well.
With the increase of $n_{2}$, the slope of $H_{1}(z)$ becomes smaller so that it is more consistent with cosmic chronometers dataset at $z=1.965$.
The evolution of the 3D space shifts from deceleration to acceleration when the redshift $z\approx0.44$ which is in line with the observed value.
We would obtain more restrictions on the number of extra dimensions if cosmic chronometers dataset are more precise.

\section{Summary and conclusion}

In conclusion, we investigate a $(n+1)$-d generalized Randall-Sundrum model with a bulk cosmological constant of order $1$ in
Planck unit (the vacuum energy caused by quantum fluctuations).
The induced cosmological constant $\Omega$ on the visible brane appears naturally in this model.
Both the visible and hidden branes are stable because their tensions are positive when $\Omega<0$.
Adopting an anisotropic metric ansatz, we obtain the positive effective cosmological constant $\Omega_{eff}$ of order $-\Omega\sim10^{-124}$
and only require a solution $\simeq50-80$.
Therefore, the cosmological constant fine tuning problem can be solved quite well.

Meanwhile, the Hubble parameters $H_{1}(z)$ as a function of redshift is consistent with most of cosmic chronometers dataset.
The evolution of the 3D space naturally shifts from deceleration to acceleration when the redshift $z\approx0.4$.
Furthermore, more precise cosmic chronometers dataset may verify the existence of extra dimensions on the visible brane
and restrict the number of the extra dimensions.
It is worth noting that we did not introduce the matter and radiation energy-momentum tensor having a perfect fluid form when we obtain the above results.
This demonstrates that the evolution of 3D scale factor (cosmic acceleration period or the like-matter-dominated period)
is intrinsically an extra-dimensional phenomenon.
This can be seen as a dynamic model of dark energy which is driven by the evolution of the extra dimensions on the brane.

\acknowledgments

We wish to acknowledge the support of the State Key Program of National Natural Science Foundation of China (under Grant No. 11535005),
the National Natural Science Foundation of China (under Grant No. 11647087),
the Natural Science Foundation of Yangzhou Polytechnic Institute (under Grant No. 201917),
and the Natural Science Foundation of Changzhou Institute of Technology (under Grant No. YN1509).



\begin{thebibliography}{99}


\bibitem{Riess} A. G. Riess et al., \emph{Observational Evidence from Supernovae for an Accelerating Universe and a Cosmological Constant}, \emph{Astron.} J. {\bf 116} (1998) 1009 [astro-ph/9805201].
\bibitem{Perlmutter} S. Perlmutter et al., \emph{Measurements of Omega and Lambda from 42 High-Redshift Supernovae}, \emph{Astrophys.} J. {\bf 517} (1999) 565 [astro-ph/9812133].
\bibitem{Bennett} C. L. Bennett et al., \emph{First Year Wilkinson Microwave Anisotropy Probe (WMAP) Observations: Preliminary Maps and Basic Results}, \emph{Astrophys. J. Suppl. Ser.} {\bf 148} (2003) 1 [astro-ph/0302207].
\bibitem{Netterfield} C. B. Netterfield et al., \emph{A measurement by BOOMERANG of multiple peaks in the angular power spectrum of the cosmic microwave background}, \emph{Astrophys.} {\bf J. 571} (2002) 604 [astro-ph/0104460].
\bibitem{Halverson} N.W. Halverson et al., \emph{DASI First Results: A Measurement of the Cosmic Microwave Background Angular Power Spectrum}, \emph{Astrophys.} {\bf J. 568} (2002) 38 [astro-ph/0104489].
\bibitem{Valent} A. G\'{o}mez-Valent and L. Amendolab, \emph{$H_0$ from cosmic chronometers and Type Ia supernovae, with Gaussian Processes and the novel Weighted Polynomial Regression method}, \emph{JCAP} {\bf 04} (2018) 051 [arXiv:1802.01505].
\bibitem{Zhang} C. Zhang, H. Zhang, S. Yuan, T.-J. Zhang and Y.-C. Sun, \emph{Four new observational H(z) data from luminous red galaxies in the Sloan Digital Sky Survey data release seven}, \emph{Res. Astron. Astrophys.} {\bf 14} (2014) 1221 [arXiv:1207.4541].
\bibitem{Verde} R. Jim\'{e}nez, L. Verde, T. Treu and D. Stern, \emph{Constraints on the equation of state of dark energy and the Hubble constant from stellar ages and the CMB}, \emph{Astrophys.} {\bf J. 593} (2003) 622 [astro-ph/0302560].
\bibitem{Simon} J. Simon, L. Verde and R. Jim\'{e}nez, \emph{Constraints on the redshift dependence of the dark energy potential}, \emph{Phys. Rev.} {\bf D 71} (2005) 123001 [astro-ph/0412269].
\bibitem{Moresco1} M. Moresco et al., \emph{Improved constraints on the expansion rate of the Universe up to z 1.1 from the spectroscopic evolution of cosmic chronometers}, \emph{JCAP} {\bf 08} (2012) 006 [arXiv:1201.3609].
\bibitem{Moresco2} M. Moresco et al., \emph{A 6\% measurement of the Hubble parameter at $z\sim0.45$: direct evidence of the epoch of cosmic re-acceleration}, \emph{JCAP} {\bf 05} (2016) 014 [arXiv:1601.01701].
\bibitem{Ratsimbazafy} A.L. Ratsimbazafy et al., \emph{Age-dating luminous red galaxies observed with the Southern African
Large Telescope}, \emph{Mon. Not. Roy. Astron. Soc.} {\bf 467} (2017) 3239 [arXiv:1702.00418].
\bibitem{Stern} D. Stern, R. Jim\'{e}nez, L. Verde, M. Kamionkowski and S.A. Stanford, \emph{Cosmic chronometers:
constraining the equation of state of dark energy. I: H(z) measurements}, \emph{JCAP} {\bf 02} (2010) 008 [arXiv:0907.3149].
\bibitem{Moresco3} M. Moresco, \emph{Raising the bar: new constraints on the Hubble parameter with cosmic
chronometers at $z\sim2$}, \emph{Mon. Not. Roy. Astron. Soc.} {\bf 450} (2015) L16 [arXiv:1503.01116].
\bibitem{Peebles} P. J. E. Peebles, B. Ratra, \emph{The Cosmological Constant and Dark Energy}, \emph{Rev. Mod. Phys.} {\bf 75} (2003) 559 [astro-ph/0207347].
\bibitem{Copeland} E. J. Copeland, M. Sami, and S. Tsujikawa, \emph{Dynamics of dark energy}, \emph{Int. J. Mod. Phys.} {\bf D 15} (2006) 1753-1936 [hep-th/0603057].
\bibitem{Amendola} L. Amendola, S. Tsujikawa, \emph{Dark energy: theory and observations}, Cambridge University Press (2010).
\bibitem{Wetterich} C. Wetterich, \emph{Cosmology and the fate of dilatation symmetry}, \emph{Nucl. Phys.} {\bf B 302} (1988) 668.
\bibitem{Zlatev} I. Zlatev, L. M. Wang, and P. J. Steinhardt, \emph{Quintessence, Cosmic Coincidence, and the Cosmological Constant}, \emph{Phys. Rev. Lett.} {\bf 82} (1999) 896 [astro-ph/9807002].
\bibitem{Caldwell} R. R. Caldwell, \emph{A Phantom Menace? Cosmological consequences of a dark energy component with super-negative equation of state}, \emph{Phys. Lett.} {\bf B 545} (2002) 23 [astro-ph/9908168].
\bibitem{Feng} B. Feng, X. L. Wang, and X. M. Zhang, \emph{Dark Energy Constraints from the Cosmic Age and Supernova}, \emph{Phys. Lett.} {\bf B 607} (2005) 35 [astro-ph/0404224].
\bibitem{Sotiriou} T. P. Sotiriou, V. Faraoni, \emph{f(R) theories of gravity}, \emph{Rev. Mod. Phys.} {\bf 82} (2010) 451 [arXiv:0805.1726].
\bibitem{Dvali} G. R. Dvali, G. Gabadadze, and M. Porrati, \emph{4D gravity on a brane in 5D Minkowski space}, \emph{Phys. Lett.} {\bf B 485} (2000) 208 [hep-th/0005016].
\bibitem{Bousso} R. Bousso and J. Polchinski, \emph{Quantization of Four-form Fluxes and Dynamical Neutralization of the Cosmological Constant}, \emph{JHEP} {\bf 0006} (2000) 006 [hep-th/0004134].
\bibitem{KK} O. Klein, \emph{Quantentheorie und f\"{u}nfdimensionale Relativit\"{a}tstheorie}, \emph{Z. Phys.} {\bf 37} (1926) 895; O. Klein, \emph{The Atomicity of Electricity as a Quantum Theory Law}, \emph{Nature} {\bf 118} (1926) 516.
\bibitem{NAH1} N. Arkani-Hamed, S. Dimopoulos, and G. Dvali, \emph{The hierarchy problem and new dimensions at a millimeter}, \emph{Phys. Lett.} {\bf B 429} (1998) 263 [hep-ph/9803315].
\bibitem{NAH2} N. Arkani-Hamed, S. Dimopoulos, and G. Dvali, \emph{Phenomenology, astrophysics, and cosmology of theories with submillimeter dimensions and TeV scale quantum gravity}, \emph{Phys. Rev.} {\bf D 59} (1999) 086004 [hep-ph/9807344].
\bibitem{RS} L. Randall and R. Sundrum, \emph{Large Mass Hierarchy from a Small Extra Dimension}, \emph{Phys. Rev. Lett.} {\bf 83} (1999) 3370 [hep-ph/9905221].
\bibitem{Das} S. Das, D. Maity, and S. Sengupta, \emph{Cosmological constant, brane tension and large hierarchy in a generalized Randall-Sundrum braneworld scenario}, \emph{JHEP} {\bf 05} (2008) 042 [arXiv:0711.1744].
\bibitem{Antoniadis} I. Antoniadis, N. Arkani-Hamed, S. Dimopoulos and G. Dvali, \emph{New dimensions at a millimeter to a Fermi and superstrings at a TeV}, \emph{Phys. Lett.} {\bf B 436} (1998) 257 [hep-ph/9804398].
\bibitem{Das1}A. Das, D. Maity, T. Paul, S. SenGupta, \emph{Bouncing cosmology from warped extra dimensional scenario}, \emph{EPJC} {\bf 77} (2017) 813 [arXiv:1706.00950].
\bibitem{Sundrum} R. Sundrum, \emph{Effective field theory for a three-brane universe}, \emph{Phys. Rev.} {\bf D 59} (1999) 085009 [hep-ph/9805471].
\bibitem{Lykken} J. Lykken, L. Randall, \emph{The Shape of Gravity}, \emph{JHEP} {\bf 06} (2000) 014 [hep-th/9908076].
\bibitem{Antoniadis1} I. Antoniadis, \emph{A possible new dimension at a few TeV}, \emph{Phys. Lett.} {\bf B 246} (1990) 377.
\bibitem{Visinelli} L. Visinelli, N. Bolis, S. Vagnozzi, \emph{Brane-world extra dimensions in light of GW170817}, \emph{Phys. Rev.} {\bf D 97} (2018) 064039 [arXiv:1711.06628].
\bibitem{Vagnozzi} S. Vagnozzi and L. Visinelli, \emph{Hunting for extra dimensions in the shadow of M87*}, \emph{Phys. Rev.} {\bf D 100} (2019) 024020 [arXiv:1905.12421].
\bibitem{Paul}T. Paul and S. SenGupta, \emph{Dynamical suppression of spacetime torsion}, \emph{EPJC} {\bf 79} (2019) 591 [arXiv:1808.00172].
\bibitem{Polchinski} J. Polchinski, \emph{String Theory. Vol. 2: Superstring theory and beyond}, Cambridge University Press (1998).
\bibitem{Koley} R. Koley, J. Mitra, and S. SenGupta, \emph{Fermion localization in a generalized Randall-Sundrum model}, \emph{Phys. Rev.} {\bf D 79} (2009) 041902(R) [arXiv:0806.0455].
\bibitem{Mitra} J. Mitra, T. Paul, S. SenGupta, \emph{Fermion localization in higher curvature and scalar-tensor theories of gravity}, \emph{EPJC} {\bf 77} (2017) 833 [arXiv:1707.06532].
\bibitem{SC2} S. Chakraborty, S. SenGupta, \emph{Effective gravitational field equations on m-brane embedded in n-dimensional bulk of Einstein and f(R) gravity}, \emph{EPJC} {\bf 75(11)} (2015) 538 [arXiv:1504.07519].
\bibitem{Banerjee} I. Banerjee, S. Chakraborty, and S. SenGupta, \emph{Radion induced inflation on nonflat brane and modulus stabilization}, \emph{Phys. Rev.} {\bf D 99} (2019) 023515 [arXiv:1806.11327].
\bibitem{SC1} S. Chakraborty and S. SenGupta, \emph{Metric factorizability and equivalence of brane world models with Brans-Dicke theory}, \emph{Phys. Rev.} {\bf D 92} (2015) 024059 [arXiv:1502.00783].
\bibitem{SC3} S. Chakraborty, and S. SenGupta, \emph{Solving higher curvature gravity theories}, \emph{EPJC} {\bf 76} (2016) 552 [arXiv:1604.05301].
\bibitem{Middleton} C. A. Middleton, and E. Stanley, \emph{Anisotropic evolution of 5D Friedmann-Robertson-Walker spacetime}, \emph{Phys. Rev.} {\bf D 84} (2011) 085013 [arXiv:1107.1828].
\bibitem{Kang} G.-Z Kang, D.-S. Zhang, L. Du, J. Xu and H.-S. Zong, \emph{Anisotropic evolution of 4-brane in a 6d generalized
Randall-Sundrum model}, \emph{Chin. Phys.} {\bf C 43(9)} (2019) 095101 [arXiv:1906.04425].






\end{thebibliography}
\end{document}